\documentclass[12pt]{iopart}

\usepackage{iopams}

\usepackage{amssymb}

\usepackage[english]{babel}
\usepackage[latin1]{inputenc}

\newcommand{\dd}{\dagger}

\newcommand{\be}[1]{\begin{equation}\label{#1}}
\newcommand{\ee}{\end{equation}}
\newcommand{\ba}[1]{\begin{eqnarray}\label{#1}}
\newcommand{\ea}{\end{eqnarray}}
\newcommand{\rf}[1]{(\ref{#1})}
\newcommand{\nn}{\nonumber}

\def\a{\alpha}
\def\b{\beta}
\def\d{\delta}
\def\g{\gamma}

\def\lb{\lambda}

\def\om{\omega}

\def\vfi{\varphi}

\newcommand{\cH}{\mathcal{H}}

\newcommand{\cU}{\mathcal{U}}

\def\ra{\rangle}
\def\la{\langle}

\newcommand{\wt}{\widetilde}
\newcommand\Sc{\mbox{Schr\"odinger }}

\def\wt{\widetilde}

\def\sech{\,\mbox{\rm sech\,}}

\newcommand{\bta}{\mathbf{H}}

\newcommand{\bu}{\mathbf{u}}

\newcommand{\bH}{\mathbf{H}}

\newcommand{\bU}{\mathbf{U}}
\newcommand{\bV}{\mathbf{V}}
\newcommand{\bW}{\mathbf{W}}

\newcommand{\tb}[1]{\textbf{#1}}

\begin{document}

\normalfont

\title[]{Supersymmetric $\eta$ operators}

\author{\large Boris F. Samsonov}

\address{Physics Department, Tomsk State University, 36 Lenin Avenue,
634050 Tomsk, Russia}

\eads{\mailto{samsonov@phys.tsu.ru}}

\begin{abstract}
Being chosen as a differential operator of a special form,
metric $\eta$ operator becomes unitary
equivalent to a one-dimensional
Hermitian Hamiltonian with a natural supersymmetric
structure.
We show that fixing the superpartner of this Hamiltonian permits to
determine both the metric operator and corresponding non-Hermitian
Hamiltonian.
 Moreover, under an additional restriction on the non-Hermitian
Hamiltonian, it becomes a superpartner of another Hermitian Hamiltonian.
\end{abstract}

\section{Introduction}

After a seminal paper by
Scholtz Geyer and Hahne \cite{SGH}, metric operator $\eta$ plays a crucial
role in the pseudo-Hermitian (quasi-Hermitian) quantum mechanics. Just
this operator is used while one redefines the inner product to bring
initially non-Hermitian operator to its Hermitian form
(for details see review papers \cite{Bender}, \cite{Mostafaz}).
Usually (see e.g. \cite{Mostafaz}) this is
 an invertible, Hermitian, positive
definite and bounded operator such that
\be{LLH}
\eta H=H^\dag\eta\,.
\ee
Here $H$ is a given non-Hermitian operator (Hamiltonian) and $H^\dag$ is
its Hermitian adjoint.
If $H$ has a discrete spectrum and admits a complete set of biorthonormal
eigenvectors then the metric operator may be presented as
\cite{Most_2814}
\[
\eta=OO^\dag\,.
\]
Note that this form is typical (see e.g. \cite{myTMP,CKS})
when a Hermitian Hamiltonian is presented in a factorized  form
in supersymmetric quantum mechanics (SUSY QM).
Its superpartner $\eta_0$ is
obtained by interchanging $O$ and $O^\dag$
\[
\eta_0=O^\dag O\,.
\]
The only difference between the SUSY intertwiner,
that we will denote $L$,
 and the operator $O$ is that $L$ is a differential operator so that the
 metric $LL^\dag$ becomes a differential operator. It is clear that
 $LL^\dag$ is Hermitian.
It is both positive definite and invertible
 if it has an empty kernel in the corresponding Hilbert space.
 This is possible if both $L$ and $L^\dag$ have empty kernels. In SUSY QM
 this corresponds to a broken supersymmetry \cite{CKS}.

 Thus the only obstacle to use the technique of SUSY QM for studying
 properties of differential metric operators is that these operators,
 being differential operators, are usually unbounded. Nevertheless, for
 instance Fityo \cite{Fityo} used a first order differential operators
 $O$ for constructing a new class of non-Hermitian Hamiltonians
 with real spectra without any discussion that the corresponding metric
 operator becomes unbounded. He also mentioned that $H$ and $H^\dag$ are
superpartners of the second order supersymmetry. Nevertheless, in contrast
to the usual second order supersymmetry \cite{myTMP}, here an additional
restriction is imposed on the intertwining operator, it should be Hermitian.
Consequences of this restriction are not analyzed in \cite{Fityo}.
Our analysis shows that corresponding first order operator $L$
(actually its complex conjugate form $L^*$) intertwines $H$ with a
Hermitian operator that we will denote $h_0$.

Any unbounded operator has a domain of definition which is a subset of the
corresponding Hilbert space. Therefore while redefining the inner product
in the spirit of paper \cite{Most_JMP-I}
\be{new_in_prod}
\langle\langle\psi_1|\psi_2\rangle\rangle_\eta=
\langle\psi_1|\eta|\psi_2\rangle\,,\quad \forall\psi_1,\psi_2\in\cH
\ee
where $\cH$ is the initial Hilbert space,
one has to replace the condition $\forall\psi_1,\psi_2\in\cH$ by
another one $\forall\psi_1,\psi_2\in D_\eta\subset\cH$ where $D_\eta$ is
the domain of definition of $\eta$ operator.
From here it follows that
there exist in $\cH$ vectors $\vfi\notin D_\eta$ which cannot be mapped into
the new Hilbert space $\cH_\eta$ defined with the help of the inner product
$\langle\langle\psi_1|\psi_2\rangle\rangle_\eta$.
This is the price for using unbounded $\eta$ operators.
From the physical viewpoint, it remains to hope that maybe states described by
such vectors cannot be realized in practice.
Moreover, this is not an obstacle for finding Hermitian Hamiltonian $h$
related to the given non-Hermitian $H$ by a similarity transformation
\cite{myPLA2010}, \cite{myJPA_2011}.
In particular, in  \cite{myJPA_2011}
scattering matrix and cross section for $h$ are calculated and their
unusual properties are discussed.

\section{Second order differential $\eta$ operator}

Let we are given a
non-Hermitian Hamiltonian
\be{H}
H=-\frac{d^2}{dx^2}+V(x)\,,\quad x\ge0
\ee
 with the domain
\be{dom_H}
\fl
D_H=\{\varphi\in L^2(0,\infty):\ \varphi''(x)-V(x)\vfi(x)\in L^2(0,\infty)\,,\
[\varphi'(x)+w(x)\vfi(x)]_{x=0}=0\}
\ee
where $V(x)$ and $w(x)$ are complex valued functions.
We find
adjoint operator $H^\dag$ as usual using the inner product un the space
$L^2(0,\infty)$,
$\la H^\dag\vfi_1|\vfi_2\ra=\la\vfi_1|H\vfi_2\ra$,
$\vfi_2\in D_H$, $\vfi_1\in D_{H^\dag}$
thus obtaining
\[
\fl
D_{H^\dag}=\{\varphi\in L^2(0,\infty):\ \!\varphi''(x)-V^*(x)\vfi(x)\in L^2(0,\infty)\,,\quad
\! [\varphi'(x)+w^*(x)\vfi(x)]_{x=0}=0\}
\]
\be{Hdag}
H^\dag=-\frac{d^2}{dx^2}+V^*(x)=H^*\,.
\ee
For simplicity we will assume that $V(x)$ is a scattering potential
 satisfying the condition
\be{V_scatter}
\int_0^\infty(1+x)|V(x)|\,dx<\infty\,.
\ee
In particular,
 we will assume that
the function $|V(x)|$ is bounded below and tends
to zero faster than any finite power of $1/x$ when $x\to\infty$
so that the operator of multiplication by the function $V(x)$,
as an operator acting in $L^2(0,\infty)$,
 becomes bounded.
 For this reason, we can simplify the domains $D_H$ and $D_{H^\dag}$
\be{DH}
\fl
D_H=\{\varphi\in L^2(0,\infty):\ \varphi''(x)\in L^2(0,\infty)\,,\quad
[\varphi'(x)+w(x)\vfi(x)]_{x=0}=0\}\,,
\ee
\[
\fl
D_{H^\dag}=\{\varphi\in L^2(0,\infty):\ \!\varphi''(x)\in L^2(0,\infty)\,,\quad
\! [\varphi'(x)+w^*(x)\vfi(x)]_{x=0}=0\}
\]

As was discussed in Introduction,
to find the metric operator for the given non-Hermitian Hamiltonian $H$,
 one has to
find a Hermitian positive definite and invertible operator $\eta$
satisfying equation \rf{LLH}.
If $\eta$ is bounded its domain is the whole Hilbert space
and no problems occur to act both by the left and by the right hand sides
of \rf{LLH} on functions belonging to $D_H$.
Unfortunately, this is not our case since
we want to consider
unbounded $\eta$ having
its own domain in $L^2(0,\infty)$.
 We find resonable to assume that
 the domain of $\eta$ coincides with that of $H$,
$D_\eta=D_H$ \rf{dom_H} (or \rf{DH}).
As we show below,
if operator $\eta$ is chosen to be a
second order differential operator
of a special form defined on this domain,
this assumption is justified
since such $\eta$ is selfadjoint.
Nevertheless, even in such a case one cannot apply
\rf{LLH} to any $\psi\in D_H$.
Indeed if $\psi\in D_\eta=D_H$
then from the left hand side of \rf{LLH} it follows that
$H\psi\in D_H$. Thus \rf{LLH} has a sense on a subset of $D_H$
such that the range of $H$ is contained in the domain of $H$.
It is possible to show that this subset is dense in
$L^2(0,\infty)$ but we do not dwell on its proof.

Assume that $\eta$ is defined
with the help of a differential expression.
Then
from the condition $D_\eta=D_H$ it follows that $\eta$
may be a
second order differential operator for which we assume the form
\be{etaLLd}
\eta=LL^\dag
\ee
where
\be{LandLd}
L=-d/dx+w^*(x)\,,\quad L^\dag=d/dx+w(x)\,.
\ee
so that
\be{eta_psi}
\fl
\eta\psi(x)=-\psi''(x)+(w^*-w)\psi'(x)+[\,ww^*-w'\,]\psi(x)\,,
\quad \psi(x)\in D_\eta=D_H\,.
\ee
We note that operators $L$ and $L^\dag$ are mutually adjoint with respect
to the inner product in $L^2(0,\infty)$.
To see that consider for instance
\[
\fl
\la\psi_1|L^\dag\psi_2\ra=
\int_0^\infty\psi_1^*(\psi_2'+w\psi_2)\,dx
=[\,\psi_1^*\psi_2\,]_0^\infty
+\int_0^\infty(-\psi_1'+w^*)^*\psi_2\,dx=\la L\psi_1|\psi_2\ra\,.
\]
The integrated term here vanishes at infinity since the functions
$\psi_{1}(x)$ and $\psi_2(x)$ are smooth enough and belong to
$L^2(0,\infty)$ and therefore they tend to $0$ as $x\to\infty$.
At $x=0$ it
vanishes since $\psi_1\in D_L$ and
the domain of $L$ coincides with the range of $L^\dag$
which according to \rf{DH} reads
\be{DL}
D_L=\{\psi\in L^2(0,\infty):\psi'(x) \in L^2(0,\infty),\ \psi(0)=0\}\,.
\ee

It is easy to check that the operator $\eta$ \rf{eta_psi}
is  selfadjoint with respect to the inner product in $L^2(0,\infty)$.
Indeed, as usual assuming that $\psi_1\in D_\eta$ and
integrating by parts twice  the term with the second derivative
and once the term with the first  derivative yields
\[
\fl
\begin{array}{lcl}
\la\psi_2|\eta\psi_1\ra
&=&  \displaystyle
\int_0^\infty\psi_2^*
\bigl[\,-\psi''_1+(w^*-w)\psi'_1+(ww^*-w')\psi_1\bigr]\,dx\\[.7em]
&=&     \displaystyle
\bigl[\,\psi_2^*{}'\psi_1-\psi_2^*\psi'_1
+\psi_2^*\psi_1(w^*-w)\,\bigr]_{x=0}^\infty
\\[.7em]
& &+       \displaystyle
\int_0^\infty
\bigl[\,-\psi_2^*{}''+(w-w^*)\psi_2^*{}'
-(w^*{}'-w')\psi_2^*+(ww^*-w')\psi_2^*\,\bigr]\,\psi_1\,dx
\\[.7em]
&=&        \displaystyle
\int_0^\infty
\bigl[\,-\psi_2''+(w^*-w)\psi_2'+(ww^*-w')\psi_2\,\bigr]^*\psi_1\,dx
\\[.7em]
&=&\la\eta\psi_2|\psi_1\ra\,.
\end{array}
\]
To justify the last equality we consider the integrated term
at $x=0$
\ba{}\nn
\fl
\left[\,\psi_2^*{}'\psi_1-\psi_2^*\psi'_1
+\psi_2^*\psi_1(w^*-w)\,\right]_{x=0}
&=&
\left[\,\psi_2^*{}'\psi_1-\psi_2^*(-w\psi_1)
+\psi_2^*\psi_1(w^*-w)\,\right]_{x=0}
\\[.5em]
\nn
&=&
\psi_1(0)\left[\,\psi_2^*{}'+\psi_2^*w^*\,\right]_{x=0}
\\[.5em]
\nn
&=&0\,.
\ea
The first line here
 follows from the property that $\psi_1\in D_\eta=D_H$
given in \rf{DH} and in
the last line we used $\psi_2\in D_\eta=D_H$.
In Section \ref{sect_example} we consider examples
illustrating selfadjointness of $\eta$
when it is a second order differential operator
with constant coefficients ($w'(x)=0$)
and with variable coefficients ($w'(x)\ne0$).

Furthermore, if we impose the condition that the operator
$L^\dag$ has the empty kernel in $D_\eta$ then $\eta$ is
positive definite.
Imposing additionally that
 the operator $L$ has the empty kernel in $D_L$ \rf{DL},
 we get an invertible operator $\eta$.
With these assumptions,
 we have operator $\eta$ suitable for constructing
operator $\rho=\eta^{1/2}$ (see below).

Another property that follows from the condition $D_\eta=D_H$
\rf{dom_H} (or \rf{DH})
is that the differential equation
\be{eta_lb}
\eta\Psi_k(x)=\lb(k)\Psi_k(x)\,,\quad k\ge0
\ee
should have
 only one singular point which is $x=\infty$. This means that the
function $w(x)$ should be regular for all $x\in(0,\infty)$
 and therefore we can put
\be{wu}
w(x)=\frac{d\log[u(x)]}{dx}\,,\quad
u(x)\ne0\quad \forall x\in(0,\infty)\,.
\ee
Since $D_\eta=D_H$ \rf{DH} we have to supply equation
\rf{eta_lb} with the following
boundary condition at $x=0$
\be{bPsieq}
[\Psi'_k(x)+w(x)\Psi_k(x)\,]_{x=0}=0\,.
\ee
For simplicity we will consider the case when operator $\eta$ has no bound
states. Therefore we will impose on the functions $\Psi_k(x)$
 an asymptotic condition at $x\to\infty$ so that they describe scattering
 states of the operator $\eta$.

\section{Unitary equivalence between $\eta$ operator and a Hamiltonian}

Let us put
\be{ux}
u(x)=\rho(x) e^{i\om(x)}
\ee
where $\rho(x)$ and $\om(x)$ are real valued functions.
Then
after a unitary transformation
\be{etaUbH}
\Psi_k=\cU\bPsi_k\,,\quad \eta=\cU\bta \cU^\dag
\ee
with the operator $\cU$ being a multiplication operator on function
(actually it is simply a phase factor)
\be{UandUm1}
\cU=\left(\frac{u^*}{u}\right)^{1/2}=e^{-i\om}\,,\quad
\cU^\dag=\cU^{-1}=\left(\frac{u}{u^*}\right)^{1/2}=e^{i\om}
\ee
equation \rf{eta_lb} reduces to a Schr\"odinger-like equation
\be{bPsieqn}
\bta\bPsi_k(x)=\lb(k)\bPsi_k(x)\,,\quad
\bta=-\frac{d^2}{dx^2}+\bV(x)
\ee
where the potential $\bV(x)$ is defined in terms of a real
valued function (superpotential)
$\bW(x)$ as follows
\be{Vbld}
\bV(x)=\bW^2(x)-\bW'(x)
\ee
and the function $\bW(x)$ is expressed in terms of the modulus of
the function $u(x)$ \rf{ux}
\be{boldW}
\bW(x)=\frac12\,\,[\,\log(uu^*)\,]'=[\,\log\rho\,]'
=\frac{\rho'(x)}{\rho(x)}=\bW^*(x)\,.
\ee
Boundary condition for equation \rf{bPsieqn} follows from that for
equation \rf{eta_lb}
\be{BCnW}
\bigl[\bPsi'_k(x)+\bW(x)\bPsi_k(x)\bigr]_{x=0}=0
\ee
and from \rf{dom_H} we get
\be{dom_BH}
\fl
D_{\bH}=\{\bPsi\in L^2(0,\infty):
\bPsi''(x)-\bV(x)\bPsi(x)\in L^2(0,\infty)\,,\
[\bPsi'(x)+\bW(x)\bPsi(x)]_{x=0}=0\}\,.
\ee
Potential \rf{Vbld} has the structure typical for the supersymmetric quantum
mechanics \cite{myTMP}.
Its SUSY partner $\bV_0$ has the form
\[
\bV_0(x)=\bW^2(x)+\bW'(x)=\frac{\rho''(x)}{\rho(x)}\,.
\]
and
for the potential
$\bV(x)$ we obtain a formula typical for the Darboux transformed
potential (see e.g. \cite{myTMP})
\be{bVx}
\bV(x)=\bV_0(x)-2\bW'(x)=\bV_0(x)-2\,[\,\log\rho(x)\,]''\,.
\ee
Moreover, the function
$\rho(x)$
is a solution to the Schr\"{o}dinger equation
with the Hamiltonian
\be{H0}
\bH_0=-\frac{d^2}{dx^2}+\bV_0(x)
\ee
 corresponding
to zero eigenvalue
\be{H0rho0}
\bH_0\rho(x)=0\,.
\ee
Function $\rho(x)$ is known as a transformation function
for the Hamiltonian $\bH_0$
(see e.g. \cite{myTMP}).
If solutions to the \Sc
 differential equation
with the potential $\bV_0(x)$
are known
\be{H0PsiE}
\bH_0\bPsi_{E}^{(0)}(x)=E\bPsi_{E}^{(0)}(x)
\ee
then solutions $\bPsi_E(x)$
to the same equation with the potential $\bV(x)$ may be
obtained with the help of the transformation operator
\be{Lrho}
L_\rho=-d/dx+\bW(x)=-d/dx+\rho'(x)/\rho(x)
\ee
as
\[
\bPsi_E(x)=L_\rho\bPsi_{E}^{(0)}(x)=
-[\bPsi_{E}^{(0)}(x)]'+\bW(x)\bPsi_{E}^{(0)}(x)\,.
\]
Hence,
Hamiltonians $\bH$ and $\bH_0$ are intertwined by the operators
$L_\rho$ and $L^\dag_\rho$
\be{LbH}
L_\rho\bH_0=\bH L_\rho\,,\quad
L_\rho^\dag\bH=\bH_0 L_\rho^\dag\,,
\ee
where
\be{Ldag}
L_\rho^\dag=d/dx+\bW(x)\,.
\ee
From the first relation \rf{LbH} follows that
$L_\rho$ transforms
eigenfunctions $\bPsi_k^{(0)}(x)$ of the Hamiltonian $\bH_0$
\rf{bPsieqn} to eigenfunctions $\bPsi_k(x)$
of the Hamiltonian $\bH$
\[
\bPsi_k(x)=\lb^{-1/2}(k)L_\rho\bPsi_k^{(0)}(x)\,.
\]
From the second relation \rf{LbH} follows that
$L_\rho^\dag$ transforms
eigenfunctions $\bPsi_k(x)$ of the Hamiltonian $\bH$
\rf{bPsieqn} to eigenfunctions $\bPsi_k^{(0)}(x)$
of the Hamiltonian $\bH_0$
\be{bPsik0}
\bPsi_k^{(0)}(x)=\lb^{-1/2}(k)L_\rho^\dag\bPsi_k(x)=
\lb^{-1/2}(k)[\,\bPsi'_k(x)+\bW(x)\bPsi_k(x)\,]\,,
\ee
\be{bH_0}
\bH_0\bPsi_k^{(0)}(x)=\lb(k)\bPsi_k^{(0)}(x)\,.
\ee
Comparing equation \rf{bPsik0}
 with \rf{BCnW}, we find boundary conditions for the
\Sc equation with the Hamiltonian $\bH_0$,
\be{bPsi00}
\bPsi_k^{(0)}(0)=0\,.
\ee
It is easy to check that function
\be{bu}
\bu(x)=\frac{1}{\rho(x)}
\ee
satisfies both the equation
\be{bHbu}
\bH\bu(x)=0
\ee
and the boundary condition \rf{BCnW}.

Operator $\bH$ is unitary equivalent to $\eta$. Therefore to fulfill the
assumption that $\eta$ has a continuous spectrum without bound states,
we have to assume
that the operator $\bH_0$ also has purely continuous spectrum and the Darboux
transformation with the transformation function $\bu(x)$ \rf{bu}
over the Hamiltonian $\bH$
does not create bound state.
Furthermore, we have to assume also that the Darboux transformation from
the Hamiltonian $\bH_0$ to the Hamiltonian $\bH$ realized with the help of
the transformation function $\rho(x)$ does not create bound state either.
Assuming additionally that the potential $\bV_0(x)$ is scattering, we have
to choose solution to equation \rf{bHbu} such that
\be{rho_assympt}
\rho\to e^{dx}\,,\quad x\to\infty\,,\quad d<0\,.
\ee
The choice $d<0$, which we accept for what follows,
guaranties an increasing asymptotic behavior
 of the function $\bu(x)$ \rf{bu}.
Therefore this function is not square integrable and $\lb(k)=0$
does not belong to the spectrum of $\bH$.
In this case operators $\eta$, $\bH$ and $\bH_0$ are isospectral
and corresponding supersymmetry is broken.
We would like to emphasize
that no any restriction on the phase $\om(x)$ of
the function $u(x)$ \rf{ux} is imposed.

\section{SUSY partner for $\eta$ operator}

Operators $L$ and $L^\dag$ \rf{LandLd}
 play a crucial role in our approach since they
define both $\eta$ operator and, as we show below, the Hamiltonians $H$
and $H^\dag$.
Operators \rf{LandLd}
 are uniquely determined when the function $u(x)$ is fixed
(see equations  \rf{LandLd} and \rf{wu}).
Therefore properties
of $\eta$ operator and Hamiltonians $H$ and $H^\dag$ depend on properties
of the function $u(x)$.
Note that the modulus of $u(x)$ is defined by the Hamiltonian $\bH_0$ as a
solution to equation \rf{H0rho0}.
Therefore both the form of the operator $\eta$
and properties of the Hamiltonian $H$
depend on properties of the Hamiltonian $\bH_0$.


Equation \rf{eta_lb} may be considered as
 equation \rf{bPsieqn}
unitary transformed
with the help of operator $\cU^{-1}$ \rf{UandUm1}
 and therefore,
as we show below,
  it inherits all supersymmetric properties of
 equation  \rf{bPsieqn}.

Applying operator ${L}^\dag$ to both sides of equation \rf{eta_lb}
yields,
\be{}
({L}^\dag L){L}^\dag\Psi_k(x)=\lb(k){L}^\dag\Psi_k(x)\,.
\ee
Above discussed condition $\lb(k)\ne0$ implies
\be{}
\Psi_k^{(0)}(x)={L}^\dag\Psi_k(x)\ne0\,.
\ee
 From here one deduces that
\be{LdLast}
\eta_0\Psi_k^{(0)}(x)=\lb(k)\Psi_k^{(0)}(x)\,,\quad\eta_0={L}^\dag L\,.
\ee
Operators $\eta$ \rf{etaLLd}
 and $\eta_0$ \rf{LdLast} are intertwined by ${L}^\dag$ and $L$,
\be{eta_intertwin}
\eta_0{L}^\dag={L}^\dag\eta\,,\quad
{L} \eta_0 =\eta{L}\,.
\ee
Thus  operator ${L}^\dag$ maps solutions of Eq. \rf{eta_lb}
(eigenfunctions of $\eta$)
 to
solutions of Eq. \rf{LdLast} (eigenfunctions of $\eta_0$
which is a SUSY partner of $\eta$).
Quite similarly, operator $L$ realizes an
inverse mapping.
The transformation function for each mapping is either
that which is annihilated by the operator $L$ or that which is
annihilated by its adjoint ${L}^\dag$.
The mapping preserving the $\d$-function normalization
of the functions $\Psi_k$ and
$\Psi_k^{(0)}$ is given by
\be{wtPsi_Psi}
\Psi_k(x)=\lb^{-1/2}(k)\,L\Psi_k^{(0)}(x)\,,\quad
\Psi_k^{(0)}(x)=\lb^{-1/2}(k)\,{L}^\dag\Psi_k(x)\,.
\ee

Note that intertwining relations \rf{eta_intertwin}
 are nothing but the associativity of operator
multiplication
\be{associat}
(L^\dag L)L^\dag=L^\dag(LL^\dag)\quad
L(L^\dag L)=(LL^\dag)L\,.
\ee
We can resolve intertwining relations \rf{LLH}
in a similar way.
For that we need not only operators $L$ and $L^\dag$ but
also their complex conjugate form
\be{Laster}
L^*=-d/dx+w(x)\,,\quad (L^*)^\dag=d/dx+w^*(x)\,.
\ee
We assume here that the operation of the complex
 conjugation commutes with
 the operation of the Hermitian conjugation.

 Let us put
 \be{HLL}
H=L^*L^\dag+\a\,,\quad H^\dag=L(L^*)^\dag+\a^*
 \ee
 where $\a$ is a complex constant.
 Then under an additional assumption
 \be{LdLa_assumption}
L^\dag L^*+\a=(L^*)^\dag L+\a^*
 \ee
we reduce equation \rf{LLH} to the identity.
We thus expressed Hamiltonians $H$ and $H^\dag$
in terms of the function $w(x)$ and, taking into account
formula \rf{wu}, in terms of the function $u(x)$.
We would like to emphasize that if the Hamiltonian $\bH_0$ is fixed
then the absolute value $\rho(x)$ of the function $u(x)$ is determined
from equation \rf{H0rho0} but its phase $\om(x)$
still remains arbitrary.
Below we show that
 this arbitrariness may be fixed with the help of
  condition \rf{LdLa_assumption}.

\section{Fixing phase $\om(x)$ with the help of a Hermitian Hamiltonian
$h_0$ \label{sect_5}}

Note that the right hand side of equation \rf{LdLa_assumption} is
Hermitian conjugate with respect to its left hand side.
This means that this equation becomes
identity if operator
\be{h0a}\fl
h_0=L^\dag L^*+\a=-\frac{d^2}{dx^2}+w^2(x)+w'(x)+\a
=-\frac{d^2}{dx^2}+\frac{u''(x)}{u(x)}+\a
\ee
is Hermitian, $h_0=h_0^\dag$. The necessary condition for that is the
reality of the function
\be{v0}\fl
v_0(x)=\frac{u''(x)}{u(x)}+\a=
\frac{\rho''(x)}{\rho(x)}-[\om'(x)]^2+\b+
i\left(\om''(x)+2\frac{\rho'(x)}{\rho(x)}\,\om'(x)+\g\right)
\ee
where we put
\be{abg}
\a=\b+i\g
\ee
 and used equation \rf{ux}.
From here we find the equation for the function $\om(x)$
\be{omx}
\om''(x)+2\frac{\rho'(x)}{\rho(x)}\,\om'(x)+\g=0
\ee
and the potential
\be{v0r}
v_0(x)=\frac{\rho''(x)}{\rho(x)}-[\om'(x)]^2+\b=v_0^*(x)
\ee
defining Hermitian Hamiltonian
\be{h_0}
h_0=-\frac{d^2}{dx^2}+v_0(x)=h_0^\dag=h_0^*\,.
\ee
From formula \rf{h0a} we extract two important consequences.
First, comparing it with \rf{HLL}, we conclude that the operators $h_0$ and
$H$ are intertwined by operator $L^*$ \rf{Laster},
\be{interh0H}
L^*h_0=HL^*\,.
\ee
Second, function $u(x)$ is an eigenfunction of $h_0$,
\be{h0ux}
h_0u(x)=\a u(x)\,.
\ee

Now we can formulate two approaches for finding a pair of operators $H$
and $\eta$, satisfying equation \rf{LLH}. Both approaches are based on the
existence of an exactly solvable Hermitian Hamiltonian.

In the first
approach it is assumed that we know solutions to equation \rf{h0ux}
with the given potential $v_0(x)$
both as a
differential equation and as a spectral problem on the space of smooth
enough functions from $L^2(0,\infty)$.
To be consistent with the previous assumptions
imposed on the  Hamiltonians $H$ and $\bH$
(see, e.g., equation \rf{V_scatter}),
here we will assume that $v_0(x)$
is a scattering potential and $h_0$
has a purely continuous spectrum
\[
h_0\psi_k(x)=k^2\psi_k(x)\,,\quad k\ge0\,.
\]
Then taking a nodeless complex valued solution to equation \rf{h0ux}
(parameter $\a$ may be both real and complex), we construct,
with the help of equations \rf{Laster} and \rf{wu},
 transformation operator $L^*$.
Moreover,
from equation \rf{interh0H} we conclude that the Hamiltonian
$H$ \rf{H} is a SUSY
partner of $h_0$ \rf{h0a} and therefore
\be{Vx}
V(x)=v_0(x)-2[\log u(x)]''\,.
\ee
 Operating with the operator $L^*$ \rf{Laster}
  on the eigenfunctions of the
 Hamiltonian $h_0$, we obtain eigenfunctions of the Hamiltonian $H$.
 Operator $\eta$ is found from equation \rf{etaLLd}.
 To find Hermitian operator $h$ equivalent to non-Hermitian $H$, we have to
 solve eigenvalue equation \rf{eta_lb} for the operator $\eta$. For this
 purpose eigenvalue equations for the Hamiltonians $\bH$ \rf{bPsieqn} or
 $\bH_0$ \rf{H0} may be useful.

Using intertwining relation \rf{interh0H}
and properties $h_0=h_0^\dag=h_0^*$ and $H^\dag=H^*$
(see equations \rf{v0r} and \rf{Hdag}) yields
 \[
h_0(L^*)^\dag=(L^*)^\dag H\,.
 \]
This relation means that operators $h_0$ and $H$ are SUSY partners and
  the operator $(L^*)^\dag$ \rf{Laster} transforms eigenfunctions
 $\vfi_k(x)$
 of the Hamiltonian $H$ to eigenfunctions $\psi_k(x)$ of the Hamiltonian
 $h_0$ \cite{myJPA_2011}
 \[
\psi_k(x)=(k^2-\a)^{1/2}[\,\vfi'_k(x)+w(x)\vfi_k(x)\,]\,.
 \]
 Comparing this equation with the boundary condition \rf{DH} for the
 eigenfunctions $\vfi_k(x)$
  of the Hamiltonian $H$, we find the boundary condition for
 the eigenfunctions $\psi_k(x)$ of the Hamiltonian $h_0$
 \[
\psi_k(0)=0\,.
 \]
As was mentioned in Introduction, we are interested to have a broken
supersymmetry.
Therefore we will consider the case when $H$ and
$h_0$ are isospectral. Hamiltonian $H$ is assumed to have only continuous
spectrum, hence, $h_0$ also should have no bound states and the
transformation function should coincide with the Jost solution for the \Sc
equation with the Hamiltonian $h_0$.
Then taking into account equation \rf{h0ux} and
condition \rf{rho_assympt}
yields the asymptotic behavior of the
transformation function $u(x)$,
\be{ux_alpha}
u(x)\to e^{(d+ib)x}\,,\quad d<0\,,\quad \a=-(d+ib)^2\,,
\ee
where $b$ is an arbitrary constant.
From here and equation \rf{abg}
 one finds the relation between constants $d$, $b$
and $\b$, $\g$ (see \rf{v0})
\be{bt_gm}
\b=b^2-d^2\,,\quad\g=-2db\,.
\ee

This approach is realized in \cite{myJPA_2011}.

 In the second approach the starting point is the spectral problem
 \rf{bH_0}, \rf{bPsi00} with the given potential $\bV_0(x)$.
 Absolute value of the transformation function is
 found from equation \rf{H0rho0} and its phase is fixed by equation
 \rf{omx}. Potential $\bV(x)$ follows from equation \rf{bVx} and
 $\eta$ operator is fixed by equations \rf{etaUbH} and \rf{UandUm1}.
 Once both the modulus and phase of the transformation function \rf{ux} are
fixed, we reconstruct Hamiltonian $H$ \rf{HLL} and if necessary
potential $v_0$ \rf{v0r} and Hamiltonian $h_0$ \rf{h_0}.
We illustrate this approach in Section \ref{sect_example} by a simple
example.


\section{Equivalent Hermitian Hamiltonian $h$}

Once eigenfunctions $\Psi_k(x)$ and eigenvalues $\lb(k)$
of the operator $\eta$ are found,
one can write down its spectral representation
\be{}
\eta=\int_0^\infty dk\lb(k)|\Psi_k\ra\la\Psi_k|\,.
\ee
It has a unique Hermitian, positive definite
and invertible square root
\be{rho}
\rho=\eta^{1/2}=\rho^\dag\,,\quad \rho>0\,.
\ee
\be{}
\rho=\int_0^\infty dk\lb^{1/2}(k)|\Psi_k\ra\la\Psi_k|\,.
\ee
Hence, from equation \rf{LLH} one finds
the Hermitian operator $h$ equivalent to $H$,
\be{h}
h=\rho H\rho^{-1}=\rho^{-1}H^\dd\rho=h^\dag
\ee
which in our case also has a purely continuous
real and non-negative spectrum.
Its eigenfunctions $\Phi_k$,
\[
h\Phi_k=k^2\Phi_k\,,\quad k\ge0\,,
\]
 are obtained by applying the operator $\rho$
to the eigenfunctions of $H$ \cite{myJPA_2011}
\be{Phik}
\Phi_k=(k^2-\a)^{-1/2}\rho\,\vfi_k=(k^2-\a)^{-1}\rho L^*\psi_k\,.
\ee
The factor $(k^2-\a)^{-1/2}$ is introduced to guarantee both the normalization of
these functions,
\be{}
\la\Phi_{k'}|\Phi_k\ra=\d(k-k')\,,
\ee
and their completeness
\be{}
\int_0^\infty dk|\Phi_k\ra\la\Phi_k|=1\,.
\ee
From Eq. \rf{Phik} it follows that the eigenfunctions $\vfi_k$ of $h_0$ and
 $\Phi_k$ of $h$ are related by an
isometric operator $\mathbf{U}$  \cite{myJPA_2011}
\be{UPhik}
\Phi_k=\mathbf{U}\psi_k
\ee
where
\be{bfU}
\mathbf U=\rho L^*(h_0-\a)^{-1}=L(L^\dag L)^{-1/2}=
L\eta_0^{-1/2}=
\bigl(\mathbf{U}^\dag\bigr)^{-1}\,.
\ee

Using the spectral representation of the operator $h$,
 one can express $h$ in terms of $\rho$, $L^*$, $(L^*)^\dd$ and the
resolvent of $h_0$  \cite{myJPA_2011}
\be{hrho}
h=\frac{\rho L^*}{\a-\a^*}\,
\left[
\a(h_0-\a)^{-1}-\a^*(h_0-\a^*)^{-1}
\right]
(L^*)^\dd\rho\,.
\ee
Note that for any $a=d+ib$ with $d<0$ the point $\a=-a^2$ does not
belong to the spectrum of $h_0$ and the operator \rf{hrho} is well
defined.
When $\a$ is real $\a=\a_r=\a_r^*$ this expression becomes undetermined.
This indeterminacy
may be resolved using the usual l'Hospitale rule,
which yields
\be{hLHospital}
h=\rho L^*h_0(h_0-\a_r)^{-2}(L^*)^\dag\rho\,.
\ee
From here we extract an important consequence.
The real character of $\a_r$ implies that $d=0$
and, hence, the point $E=k^2=\a_r=b^2$
belongs to the continuous spectrum of $H$ and
corresponds to the spectral singularity  \cite{myJPA_2011}.
This point
belongs to the spectrum of $h_0$ also.
For this reason the resolvent of $h$
diverges at $k^2=\a_r=b^2$
and operator \rf{hLHospital} becomes undefined in the Hilbert space.
This means that the non-Hermitian
Hamiltonian $H$ with a spectral singularity
 does not have Hermitian counterpart.

Another possibility would be $b=0$, $d\ne0$. In this case $\a=-d^2$,
operator $H$ becomes Hermitian and $h$ becomes unitary equivalent to $H$.

Operator $\bU$ \rf{bfU}
 mapping eigenfunctions of $h_0$ to that of $h$ \rf{UPhik} may be written
 in terms of the eigenfunctions $\Psi_k^{(0)}$ of equation \rf{LdLast}
 \be{bUwtPsi}
\bU=L^*\int_{0}^\infty dk\, \lb^{-1/2}(k)\,\,
|\Psi_k^{(0)}\ra\la\Psi_k^{(0)}|\,.
 \ee
Although equation \rf{Phik}
(or equations \rf{UPhik}, \rf{bUwtPsi})
 formally solves the problem of finding the eigenfunctions of
$h$, it contains the non-local operator
$\rho$ (or $(L^\dag L)^{-1/2}$)
and, therefore, in general, no explicit expression for $\Phi_k$ exists.

\section{Examples \label{sect_example}}

As discussed in Section \ref{sect_5},
there exist two approaches for finding a pair of operators $H$
and $\eta$ satisfying equation \rf{LLH} and below we give two
corresponding examples. First we exemplify the second approach
since it leads to the simplest form of $\eta$ being a second
order differential operator with constant coefficients
and with the
non-Hermitian Hamiltonian $H$ first studied by {Schwartz}
\cite{Schwartz}.
Then we give an example where $\eta$ is a second order
differential operator with variable coefficient.

\subsection{$\eta$ is a second
order differential operator with constant coefficients}

Let us fix real constants $d<0$ and $b$ (see \rf{ux_alpha}).
In the simplest case
potential $\bV_0(x)$ may be a nonnegative constant
$\bV_0(x)=d^2$
so that the solution of
the spectral problem \rf{bH_0}, \rf{bPsi00}
reads
\[
\bPsi_k^{(0)}(x)=\sqrt{\frac2\pi}\,\sin(kx)\,,
\quad\lb(k)=k^2+d^2\,,
\quad k\ge0\,.
\]
Since $\bH_0=-d^2/dx^2+d^2$, from equation \rf{H0rho0} we find the absolute
value of the transformation function
\[
\rho(x)=e^{dx}\,.
\]
Next we solve equation \rf{omx} for $\om(x)$ with $\g$ as given in
\rf{bt_gm}
\[
\om(x)=-\frac{\g}{2d}\,x= bx\,,\quad b=-\frac{\g}{2d}\,.
\]
(integration constant is without importance here)
thus reconstructing the transformation function \rf{ux}
\[
u(x)=e^{(d+ib)x}\,.
\]
According to \rf{boldW} function $\bW(x)$ is constant $\bW(x)=d$.
Therefore potential $\bV(x)$ \rf{bVx} coincides with $\bV_0(x)=d^2$.

Transformation operator $L_\rho$ \rf{Lrho} is given by
\[
L_\rho=-d/dx+d
\]
with the help of which we find eigenfunctions of $\bH$ \rf{bPsieqn}
\[
\bPsi_k(x)=L_\rho\bPsi_k^{(0)}(x)=
\sqrt{\frac2\pi}\,(k^2+d^2)^{-1/2}\,[\,d\sin(kx)-k\cos(kx)\,]\,.
\]
Using unitary operator
\[
\cU=e^{-ibx}
\]
we obtain eigenfunctions
\[
\Psi_k(x)=\cU\bPsi_k(x)=
\sqrt{\frac2\pi}\,(k^2+d^2)^{-1/2}\,e^{-ibx}\,
[\,d\sin(kx)-k\cos(kx)\,]\,.
\]
of operator
\[\fl
\eta=LL^\dag=\Bigl(-\frac{d}{dx}+d-ib\Bigr)
\Bigl(\frac{d}{dx}+d+ib\Bigr)=
-\frac{d^2}{dx^2}-2ib\frac{d}{dx}+b^2+d^2\,.
\]
Using equations \rf{HLL} and \rf{ux_alpha} yields corresponding
non-Hermitian Hamiltonian
\[
H=-\frac{d^2}{dx^2}\,.
\]
Its Hermitian SUSY partner follows from \rf{h_0}, \rf{v0r} and \rf{bt_gm}
\[
h_0=-\frac{d^2}{dx^2}\,.
\]
Note that although both $H$ and $h_0$ are defined by the same differential
expression, corresponding spectral problems are different. For the
Hamiltonian $H$ we impose the boundary condition
\be{BCH}
\vfi'_k(0)+(d+ib)\vfi_k(0)=0
\ee
whereas for the Hamiltonian $h_0$ the boundary
condition reads
$\psi_k(0)=0$.
Another remark we would like to make is that the spectral problem for $H$
was first studied by Schwartz \cite{Schwartz} as one of the simplest
 problems where the spectral singularity occurs in the continuous
 spectrum of $H$ if $d=0$.


 \subsection{$\eta$ is a second order
  differential operator with variable coefficients}

According to Section  \ref{sect_5}, one can start with any
scattering Hamiltonian $h_0$ with a real-valued potential
$v_0(x)$. The main element of the whole construction is the
transformation operator $L^*$ which intertwines $h_0$ and a
non-Hermitian Hamiltonian $H$ \rf{interh0H}.

Let us choose
\[
v_0(x)=-\frac{2a^2}{\cosh^2(ax+c)}\,,\quad a>0\,,\quad c>0\,.
\]
For $c>0$
the Hamiltonian $h_0=-d^2/dx^2+v_0(x)$ has no bound states in
$L^2(0,\infty)$.
A solution of the \Sc equation with the Hamiltonian $h_0$ having
the asymptotic behavior as given in \rf{ux_alpha} reads
\[
u(x)=\frac{e^{(d+i b)x}}{a-d-ib}\,\bigl[\,a\tanh(ax+c)-d-ib\,\bigr]\,.
\]
Using \rf{wu} and this transformation function, one finds function
\[
w(x)=d+ib+\frac{2a^2}{a\sinh(2ax+2c)-2(d+ib)\cosh^2(ax+c)}
\]
which according to \rf{eta_psi} and \rf{Vx}
defines both $\eta$ operator and the non-Hermitian Hamiltonian $H$
respectively.

To illustrate selfadjointness of $\eta$ on $D_H$ \rf{dom_H} we find
operator $\bH$  \rf{bPsieqn} unitary equivalent to
$\eta$.
For that using \rf{boldW}, we first calculate
$\bW(x)$
\[
\bW(x)=d+\frac{a^2\sech(ax+c)\bigl[\,a\tanh(ax+c)-d\,\bigr]}%
{b^2+d^2+a\tanh(ax+c)\bigl[\,a\tanh(ax+c)-2d\,\bigr]}\,.
\]
Then using \rf{Vbld}, we find
 the potential
\be{Vbx}
\bV(x)=
d^2+\frac{4a^2(a^2-d^2)}{\wt{\bW}(x)}-\frac{12a^4b^2}{\wt{\bW}{^2}(x)}
\ee
where
\[
\fl
\wt{\bW}(x)=b^2+d^2-a^2+(a^2+b^2+d^2)\cosh(2ax+2c)-2ad\sinh(2ax+2c)\,.
\]
It is not difficult to see that $\bV(x)\to d^2$ as $x\to\infty$
by an exponential rule $\sim\exp(-2ax)$  and, hence, as
expected, operator $\bH$ is scattering and positive definite.

Another important point is that $\bV(x)$ is
continuous and bounded below. In
this case, according to a known result \cite{CS}, Operator $\bH$
initially defined on a set of finite and twice continuously
differentiable functions $y(x)$
satisfying boundary condition
$y'(0)\cos\phi+y(0)\sin\phi=0$
has a closed selfadjoint extension $\overline{\bH}$.
Its domain $D_{\overline{\bH}}$ where
$\overline{\bH}=\overline{\bH}{}^\dag$ is described by
the following properties \cite{CS}. If $z(x)\in D_{\overline{\bH}}$ then
\begin{enumerate}
\renewcommand{\labelenumi}{\theenumi}
\renewcommand{\theenumi}{(\alph{enumi})}
\item
$z(x)\in L^2(0,\infty)$,
\item
\label{point_ii}
$z(x)$ is continuous and has an absolutely continuous derivative
in any finite interval belonging $(0,\infty)$,
\item
$-z''(x)+\bV(x) z(x)\in L^2(0,\infty)$,
\item
$z'(0)\cos\phi+z(0)\sin\phi=0$, $\phi\in\Bbb R$\,.
\end{enumerate}
Evidently this domain differs from that described in \rf{dom_BH}
only by the condition \ref{point_ii} which we everywhere skipped for
simplicity. Going back from $\bH$ to its unitary equivalent
operator $\eta$, we conclude that $\eta=\eta^\dag$ on
$D_{\eta}=D_H$ \rf{dom_H}.

 \section{Conclusion}

 In this paper basing on the
 factorization property of the metric $\eta$ operator \cite{Most_2814}, we
 proposed a special differential form for this operator. It happens that
 such a metric operator is unitary equivalent to
 the usual one dimensional Hamiltonian
 and therefore it has a natural supersymmetric structure.
 Once the metric operator is fixed, one can reconstruct the corresponding
 non-Hermitian Hamiltonian.
This opens a way for starting not from a non-Hermitian Hamiltonian and
looking for the corresponding metric operator, but for going in the
opposite
direction, i.e., for starting with a given metric operator and presenting
the corresponding non-Hermitian Hamiltonian. An advantage of this approach
is the possibility to use the technique of SUSY QM for studying
properties of the metric operator. In particular, as a possible line of
future investigation, we are planning to use shape invariant metric
operators and study how this property is reflected by corresponding
non-Hermitian Hamiltonians.
We pointed out that such metric operators are unbounded. Therefore some
vectors from the initial Hilbert space are lost. They cannot be mapped
into the new Hilbert space
where the non-Hermitian Hamiltonian becomes Hermitian.

We illustrated our approach by two examples.
In the first example the metric
operator is a second order differential operator with constant
coefficients and $w'(x)=0$.
It happens that
in this case the corresponding non-Hermitian Hamiltonian
coincides with an operator proposed by Schwartz
\cite{Schwartz} (see also \cite{Gus})
 for illustrating the existence of the spectral singularity
 in the continuous part of the
 spectrum of this operator.
In the second example we considered
a differential
operator with variable coefficients as
the metric operator.
Using operator $\bH$ \rf{bPsieqn} unitary equivalent to
$\eta$, we have shown that $\eta$ is selfadjoint on $D_H$
\rf{dom_H}.

\end{document}